\documentclass[3p,times,procedia]{elsarticle}
\usepackage{nupha_ecrc}
\usepackage{wrapfig}
\volume{00}

\firstpage{1}

\journalname{Nuclear Physics A}

\runauth{}

\jid{nupha}

\jnltitlelogo{Nuclear Physics A}

\usepackage{amssymb}
 \usepackage{lineno}

\usepackage[figuresright]{rotating}

\newcommand{\pion}{\pi^0}
\newcommand{\h}{\textrm{h}^\pm}
\newcommand{\pout}{p_{\rm out}}
\newcommand{\rmspout}{\sqrt{\langle p_{\rm out}^{2}\rangle}}
\newcommand{\snn}{\sqrt{s_{_{NN}}}}
\newcommand{\pttrig}{p_{T}^{\rm trig}}
\newcommand{\dphi}{\Delta\phi}

\begin{document}

\begin{frontmatter}


%% \title{Title\tnoteref{label}}

%% Instructions from Editor: Please use the following \dochead only in the preprint version (e-print arXiv etc.); 
%% use empty \dochead{} when submitting to Nuclear Physics A!
\dochead{XXVIth International Conference on Ultrarelativistic Nucleus-Nucleus Collisions\\ (Quark Matter 2017)}
%% e.g. \dochead{17th International Conference on Dynamical Processes in Excited States of Solids}

\title{Study of cold and hot nuclear matter effects on jets with direct photon triggered correlations from PHENIX}


\author{J. D. Osborn for the PHENIX Collaboration}

\address{University of Michigan Physics Department, Ann Arbor, MI 48109}

\begin{abstract}
Direct photons, being colorless objects, provide an unmodified control particle that can be used in conjunction with jets to probe the quark-gluon plasma. To leading order the direct photon momentum balances the momentum of opposing jets and can therefore provide a clean handle on the jet energy. Therefore, angular correlations with direct photons provide a mechanism to study the fragmentation of the opposing jet without performing jet reconstruction. Jet fragmentation modification has been measured previously in PHENIX in central Au+Au collisions. Recent RHIC runs offer the potential to study these observables in heavy ion collisions with greater statistics and over different collision systems including asymmetric collision geometries. In this talk we present results of isolated direct photon-triggered correlations in d+Au collisions and discuss the constraints of cold nuclear matter effects on the fragmentation functions. We also present the latest results with higher statistics on direct photon-triggered correlations in Au+Au collisions including differential measurements of fragmentation function modification. Finally, we present the status of the centrality and collision species dependence of these observables, including comparisons to related dihadron correlations. Together these results can give a view of jet modification going from small to large system size.

\end{abstract}


\end{frontmatter}

 %\linenumbers

\section{Introduction}\label{intro}

Direct photons have long been considered the ``golden channel" in hadronic and heavy ion collisions because, at leading order (LO), they emerge directly from the hard scattering. Therefore, direct photons are one of the most precise measures of the initial partonic hard scattering kinematics in hadronic collisions. Additionally, since they only interact electromagnetically, the direct photon does not suffer from final-state QCD interactions. This allows the photon to probe effects from partonic dynamics before effects from gluon radiation, non-Abelian effects from color flow, or in the case of heavy ion collisions, medium interactions with the quark-gluon plasma. Additionally, the away-side scattered parton can then be modified by any strong interactions with participants of the interaction; thus the photon gives an excellent benchmark to compare and understand these interactions. \par

The PHENIX experiment has recently measured direct photon-hadron angular correlations in $p$$+$$p$, $p$+Al,$p$+Au, d+Au, and Au+Au collisions systems, highlighting the versatility of the Relativistic Heavy Ion Collider (RHIC) accelerator facility. The PHENIX detector is capable of measuring both direct photon-hadron and $\pion$-hadron correlations via its two central arms, covering an azimuthal acceptance of $\sim\pi$ radians and pseudorapidity $|\eta|<$0.35. \par

\section{Results in $p$+$p$}\label{pp}

Recent interest in multidimensional nucleon structure has led to novel predictions from QCD as a non-Abelian gauge theory~\cite{spin_struct}. To study one such prediction, direct photon-hadron and dihadron correlations were measured in $p$$+$$p$ collisions at $\sqrt{s}=$510 GeV. In a transverse-momentum-dependent (TMD) framework, where a hard and soft transverse momentum scale are measured, QCD factorization breaking has been predicted~\cite{ted_fact_break}. In hadronic collisions where at least one final-state hadron is measured, soft gluons may be exchanged in both the initial and final states, leading to the prediction that the nonperturbative parton distribution functions from individual protons no longer factorize from one another in a cross section calculation. Nearly back-to-back angular correlations are sensitive to initial- and final-state transverse momentum $k_T$ and $j_T$, so when the produced particles have large transverse momentum a TMD framework is valid.  \par

\begin{figure}[thb]
	\includegraphics[width=0.5\textwidth]{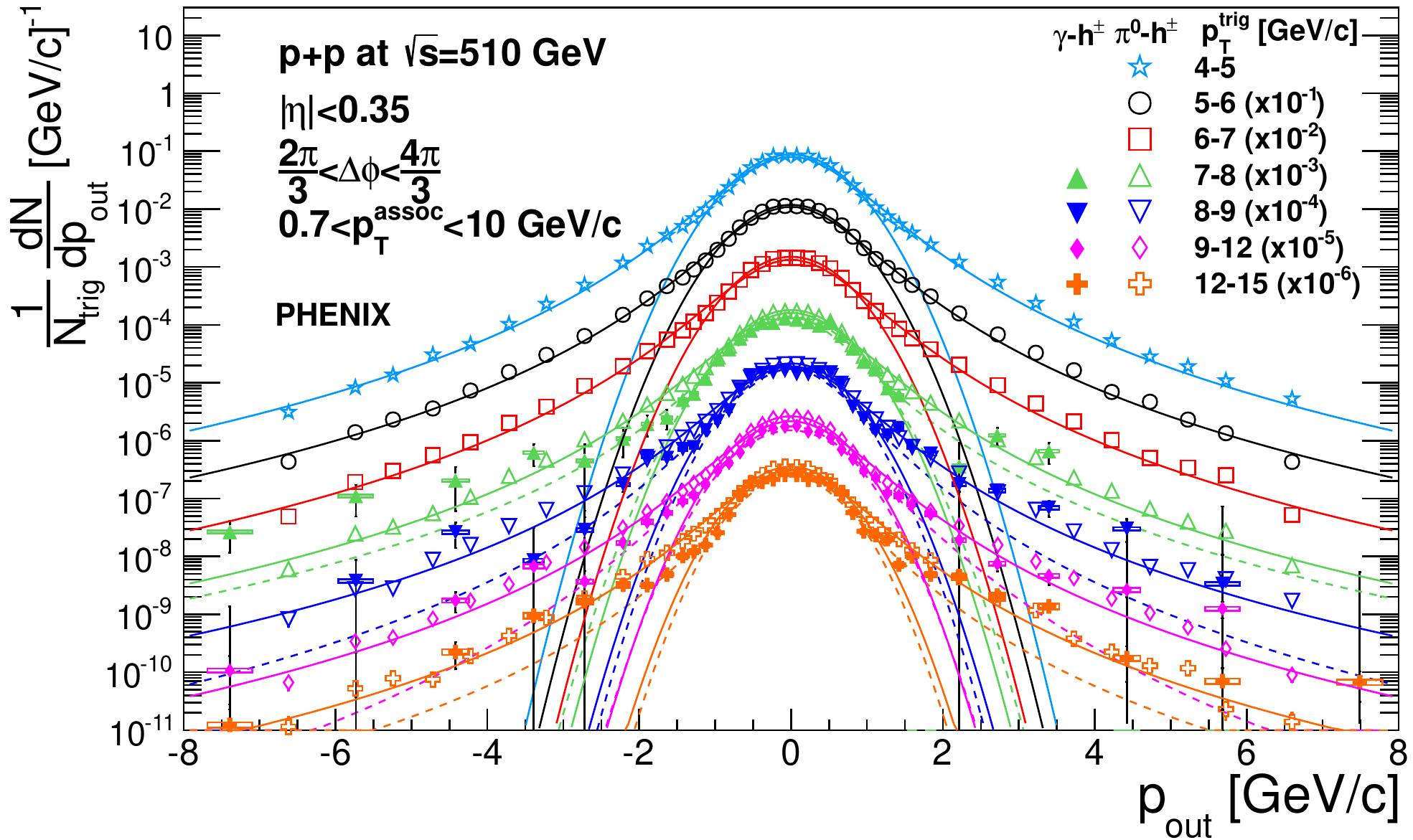}
	\includegraphics[width=0.5\textwidth]{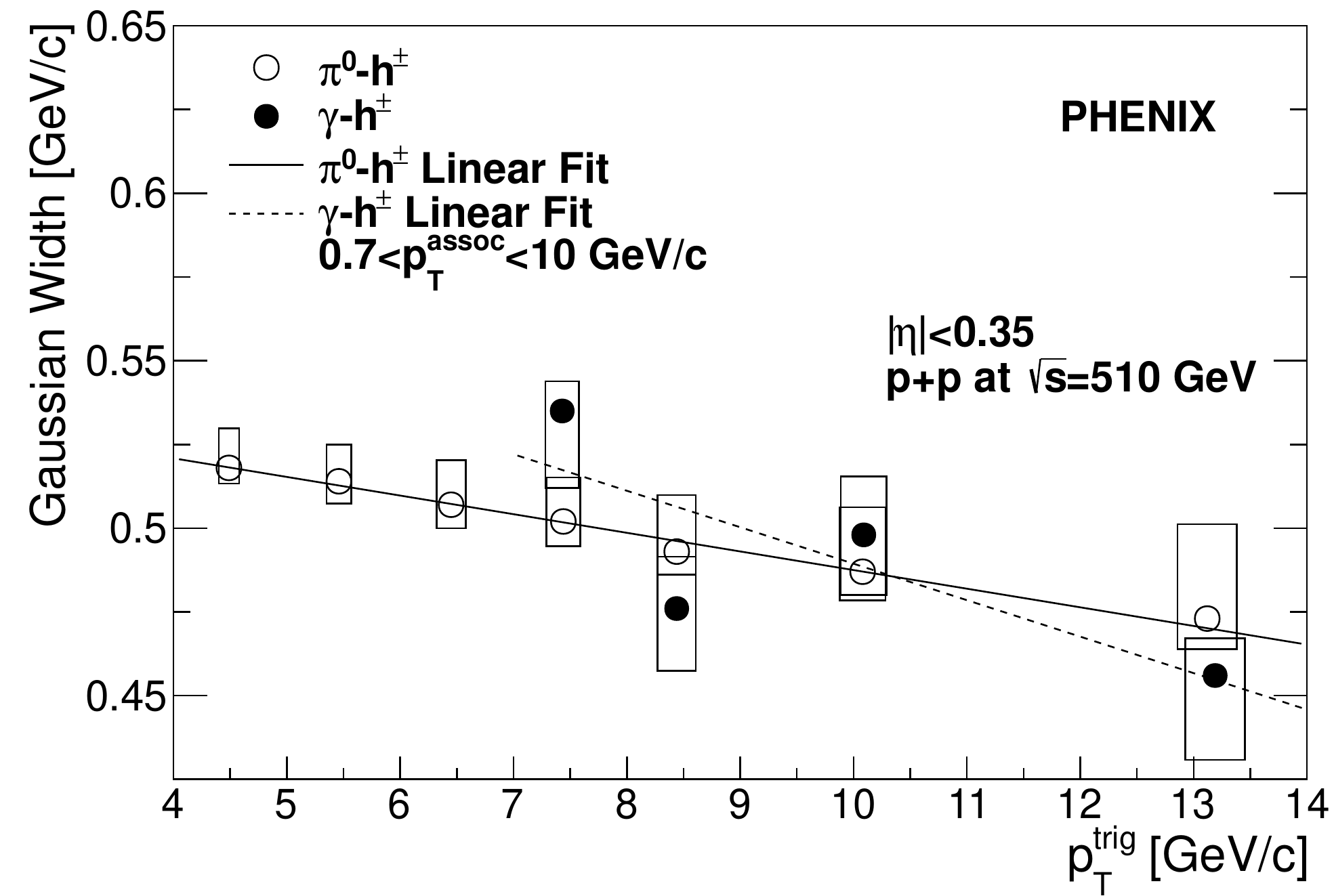}
	\caption{The measured $\pout$ distributions in $p$$+$$p$ collisions are shown for dihadron and direct photon-hadron correlations (left). The Gaussian widths of the nonperturbative $\pout$ region show a negative dependence with the hard scale of the interaction $\pttrig$ (right).}
	\label{fig:pouts}
\end{figure}

To study these effects, PHENIX measured the $\pout=p_{T}^{\rm assoc}\sin\Delta\phi$ distributions, shown in the left panel of Fig.~\ref{fig:pouts}. The observable $\pout$ is defined as the transverse momentum component perpendicular to the trigger particle, defined here as the near-side direct photon or $\pion$. The distributions show Gaussian behavior at small $\pout$ and power law behavior at large $\pout$, indicating a transition from sensitivity to nonperturbative gluon radiation to perturbative gluon radiation. Perturbative TMD evolution, which comes directly from the generalized TMD QCD factorization theorem, predicts increasing nonperturbative momentum widths with the hard scale of the interaction~\cite{john_arxiv}. Gaussian fits to the nonperturbative region of the $\pout$ distributions show that the measured momentum widths decrease with the hard scale, shown on the right of Fig.~\ref{fig:pouts}, in interactions where factorization breaking is predicted~\cite{ppg195}. \par

\section{Results in $p$+Au} \label{pA}
\begin{figure}[thb]
	
	\hspace*{-0.5cm}\includegraphics[width=0.6\textwidth]{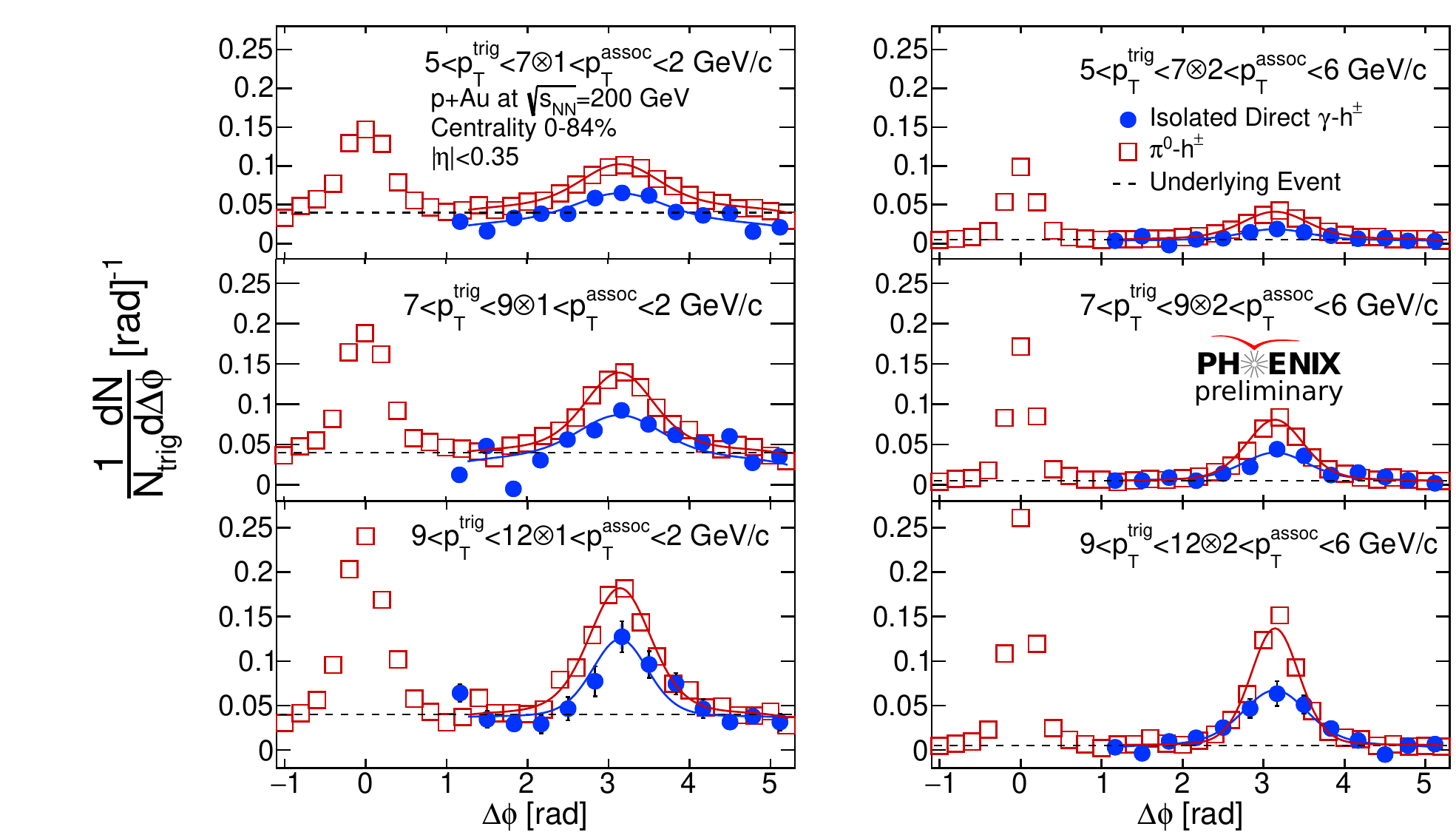}\hspace*{0.5cm}
	\includegraphics[width=0.4\textwidth]{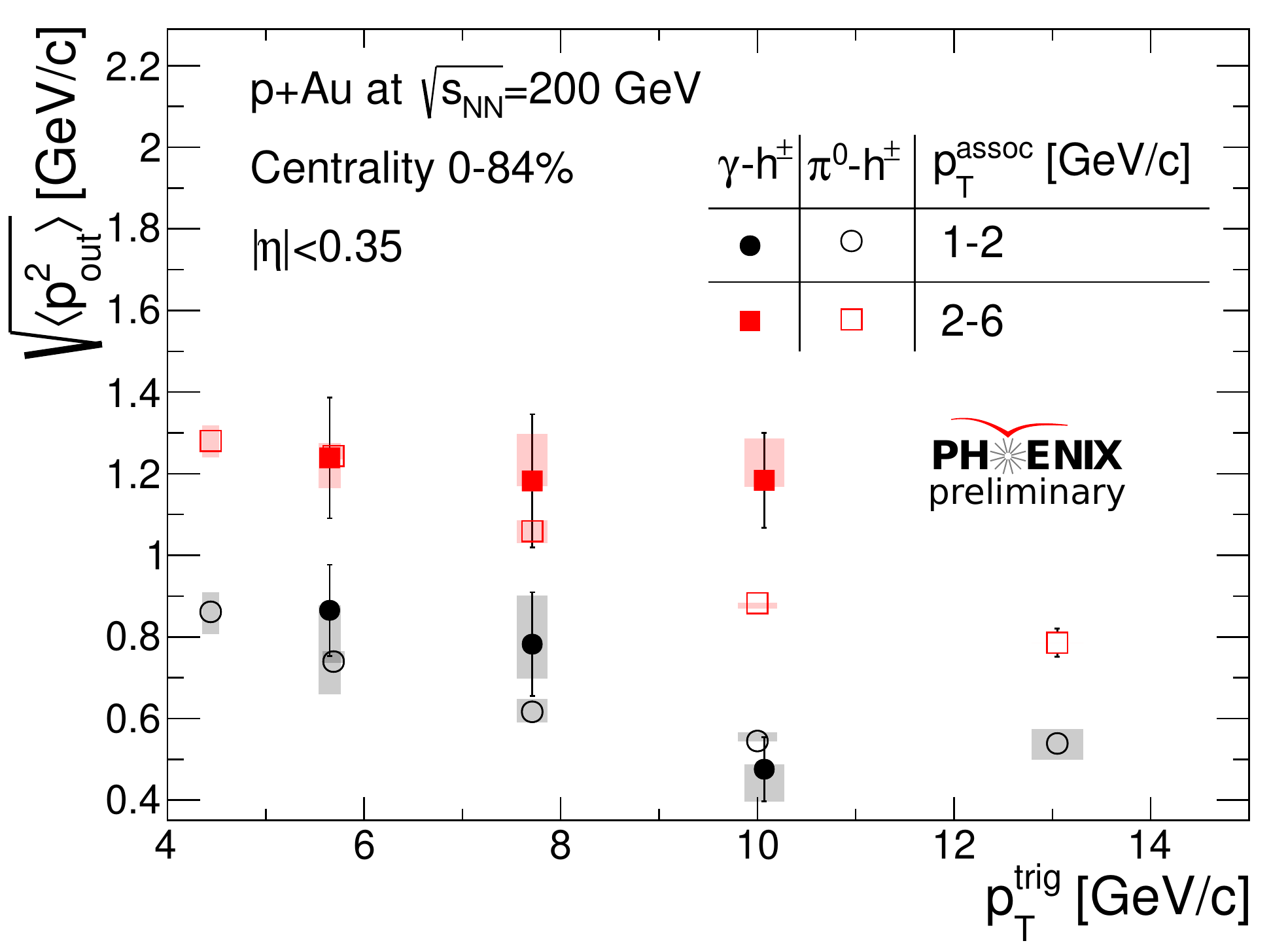}
	\caption{The measured $\dphi$ correlations in $p$+Au collisions are shown for dihadron and direct photon-hadron correlations (left). The nonperturbative momentum width $\rmspout$ is extracted from away-side fits and shown as a function of $\pttrig$ (right).}
	\label{fig:paudphis}
\end{figure}

In 2015 RHIC delivered, for the first time, $p$+Au and $p$+Al collisions at $\snn=$200 GeV. With $p$+A collisions, possible effects from factorization breaking can also be studied in a nuclear environment where soft gluon exchanges should be more abundant. Additionally, other effects could be present due to the large nucleus. The correlations as a function of $\Delta\phi$ are shown in Fig.~\ref{fig:paudphis}. The $\pion$-$\h$ correlations show a standard two-jet structure with peaks centered at $\dphi\sim0$ and $\dphi\sim\pi$. The $\gamma$-$\h$ correlations are not shown on the near-side due to the implementation of an isolation cut, which reduces the contribution from next to leading order fragmentation photons. Additionally the $\gamma-\h$ correlations show a smaller away-side yield than the $\pion-\h$ yields since they probe a smaller jet energy; the direct photons emerge from the hard scattering at LO while the $\pion$ results from a fragmented parton. The correlations are fit with a function described in Ref.~\cite{ppg195} on the away-side to extract the quantity $\rmspout$, a quantity similar to the Gaussian width of the $\pout$ distributions shown in Fig.~\ref{fig:pouts}. The results are shown in the right panel of Fig.~\ref{fig:paudphis}. When comparing to the analogous bins from Ref.~\cite{ppg195}, the $p$+Au $\rmspout$ nonperturbative momentum widths show a stronger dependence on $\pttrig$, indicating that if the decreasing widths are due to factorization breaking these effects could be stronger in $p$+A collisions. \par

\begin{wrapfigure}{r}{0.5\textwidth}
	\includegraphics[width=0.5\textwidth]{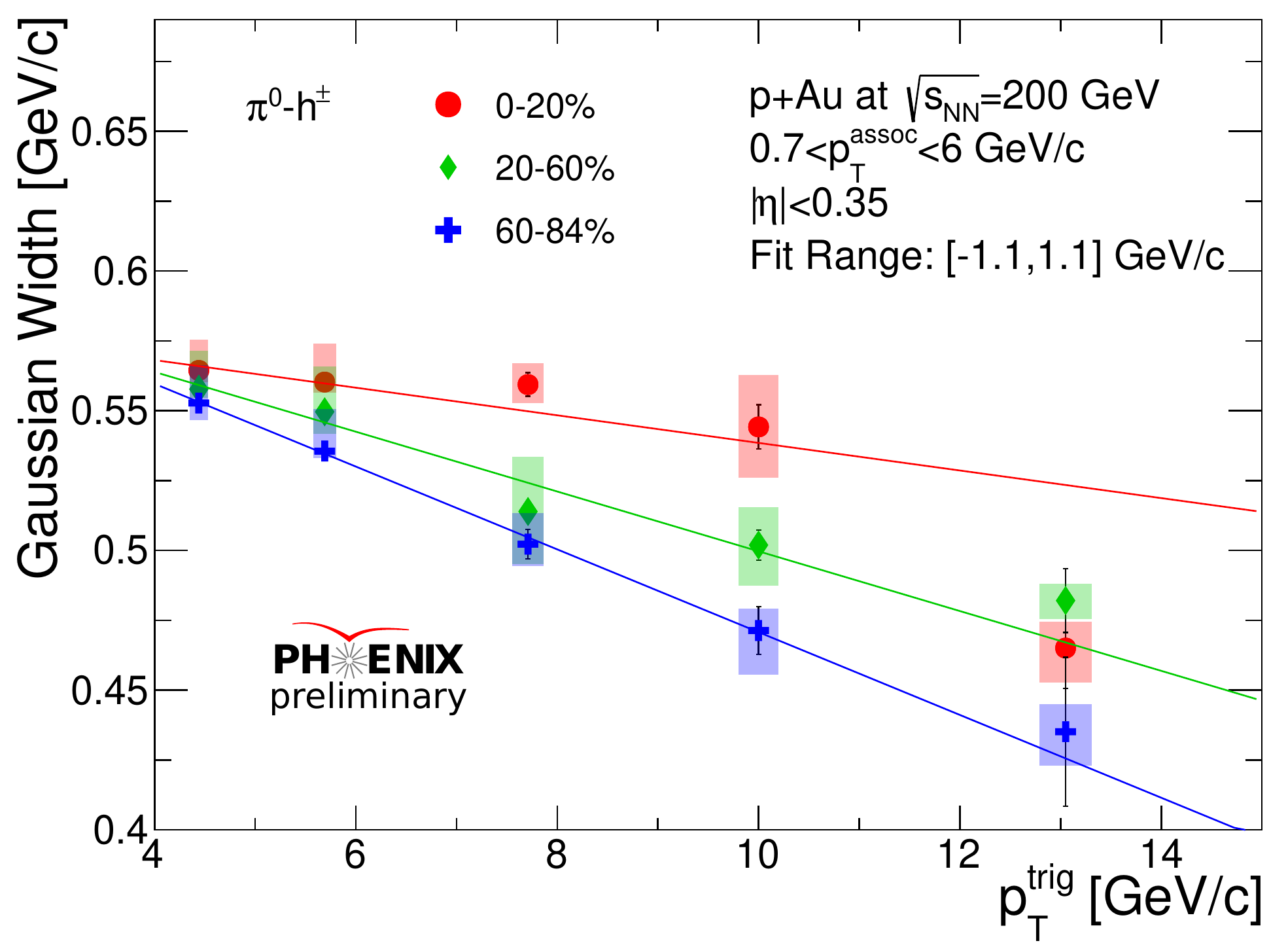}
		\caption{The dihadron nonperturbative momentum widths of $\pout$ show a clear centrality dependence in $p$+Au collisions.}
	\label{fig:centdep}
\end{wrapfigure}

In addition to possible effects from factorization breaking, the correlations were studied as a function of centrality to investigate possible effects from $k_T$ broadening and multiple scattering in $p$+A collisions. Figure~\ref{fig:centdep} shows the $\pion-\h$ Gaussian widths of $\pout$ as a function of both $\pttrig$ and centrality of the collision. The nonperturbative momentum widths show a clear centrality dependence, with more central events showing a broader distribution than more peripheral events. The centrality dependence indicates that effects from $k_T$ broadening, multiple scattering, or flow could be contributing to the broadened momentum widths in $p$+Au collisions. Although not explicitly shown in Fig. 3, a similar centrality dependence was seen in $p$+Al. \par
%This effect is additionally seen in $p$+Al in the right panel of Fig.~\ref{fig:centdep}, although the dependence on centrality is not as strong. Additionally, this shows that there is a small nuclear size dependence on the nonperturbative momentum widths as the $p$+Al widths in the same centrality class are systematically smaller than the $p$+Au widths.

\section{Results in d+Au and Au+Au}\label{AA}

Direct photon-hadron correlations can also be used to probe fragmentation functions in heavy ion collisions. Any modification to the fragmentation function in collisions of ions can be interpreted from the integrated away-side yield. At LO, since $p_T^\gamma\approx p_T^{\rm jet}$, $z_T=p_T^h/p_T^\gamma$. Therefore the fragmentation function can be written approximately as $D_q(\xi)=1/N_{evt}dN(\xi)/d\xi$, where $\xi=\ln(1/z_T)$. New direct photon-hadron results in d+Au and Au+Au investigated any fragmentation function modification by measuring the ratio $I_{AA}$, which is the ratio of the integrated away-side yield in Au+Au or d+Au collisions to $p$$+$$p$ collisions at the same energy. \par

\begin{figure}[thb]
	\includegraphics[width=0.5\textwidth]{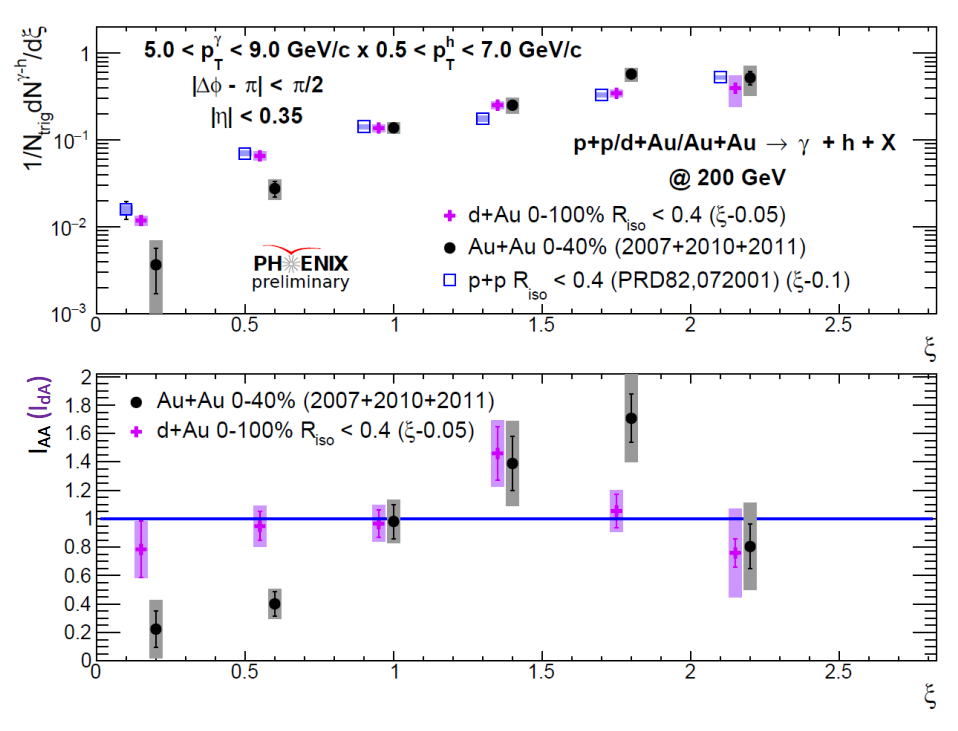}
	\includegraphics[width=0.5\textwidth]{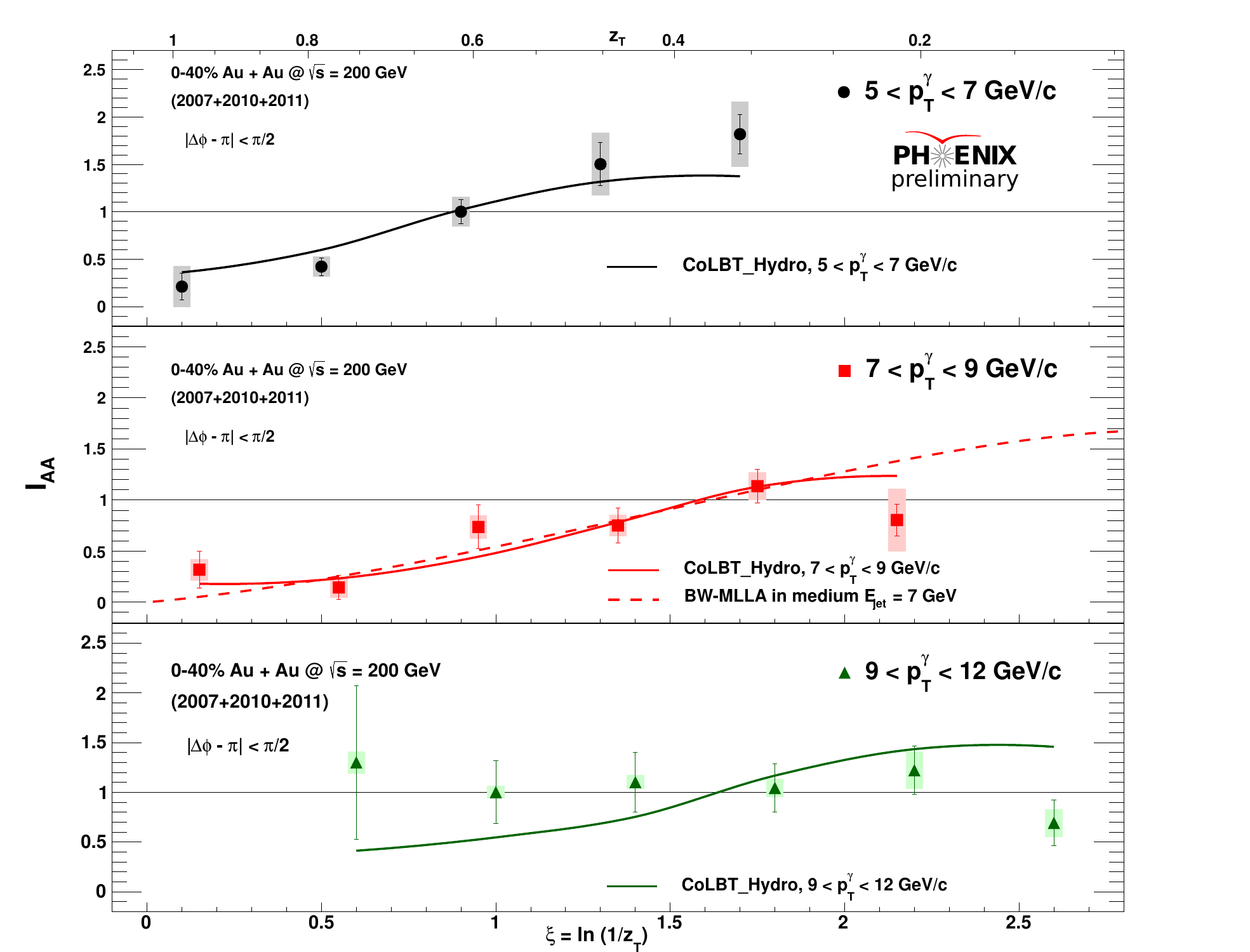}
	\caption{The integrated away-side yield $I_{AA}$ of Au+Au or d+Au to $p$$+$$p$ is shown as a function of $\xi=\ln(1/z_T)$ (left). The d+Au $I_{AA}$ results are consistent with unity, while the Au+Au results show suppression at small $\xi$ and enhancement at larger $\xi$. The Au+Au $I_{AA}$ is shown in several $\pttrig$ bins (right). The transition from suppression to enhancement occurs at different $\xi$ with increasing $\pttrig$. }
	\label{fig:auau}
\end{figure}

The measured yields differential in $\xi$ and the resulting ratio $I_{AA}$ for both d+Au and Au+Au collisions are shown in the left panel of Fig.~\ref{fig:auau}. Within uncertainties, the d+Au $I_{AA}$ values are consistent with unity, indicating that there is little to no fragmentation function modification in d+Au. The Au+Au $I_{AA}$ values show a clear difference from unity; there is suppression at small $\xi$ and an enhancement in yield at large $\xi$. Due to the increased statistical precision of the new Au+Au data set from Ref.~\cite{ppg113}, the dependence of the suppression and enhancement of $I_{AA}$ with $\pttrig$ could be investigated. This dependence is shown in three $\pttrig$ ranges in the right panel of Fig.~\ref{fig:auau}, with theory curves from Refs.~\cite{curv1, curv2}. The data show that the transition from suppression to enhancement is not at a fixed fragmentation function variable~\cite{Huijun_HP}. \par

\section{Summary}

In conclusion, new direct photon-hadron and $\pion$-hadron results from the PHENIX experiment were presented. Measurements from $p$$+$$p$, $p$+A, d+Au, and Au+Au collisions were made to study various effects in QCD systems. In $p$$+$$p$, possible effects from factorization breaking due to the non-Abelian nature of QCD were analyzed. In $p$+A collisions, these effects were investigated in a nuclear environment. A significant centrality dependence to the away-side jet widths was observed in $p$+Au collisions, indicating that effects from $k_T$ broadening or multiple scattering could be present. No fragmentation function modification was found in d+Au collisions, while significant modification was observed as a function of $p_T^\gamma$ in Au+Au collisions. Upcoming results from PHENIX with the higher statistics $p$$+$$p$, Cu+Au, and Au+Au data sets will improve the precision of the transition region of $\xi$ and will additionally allow system size dependent studies. \par



%% References


\bibliographystyle{elsarticle-num}
\bibliography{QM17_bib}






% \end{thebibliography}

\end{document}